\documentstyle[a4,12pt,epsf]{article}
\newcommand{\nn}{\nonumber}
\newcommand{\be}{\begin{equation}}
\newcommand{\ee}{\end{equation}}
\newcommand{\bea}{\begin{eqnarray}}
\newcommand{\eea}{\end{eqnarray}}
\newcommand{\bean}{\begin{eqnarray*}}
\newcommand{\eean}{\end{eqnarray*}}
\def\leftB{\Bigl}
\def\rightB{\Bigr}
\def\L{{\rm L}}
\def\R{{\rm R}}
\def\VEV#1{\langle #1 \rangle}

\def\fun#1{\!\left(#1\right)}   % for an argument of a function
\def\widebar#1{\vbox{\ialign{##\crcr%
        \hskip 1.0pt\hrulefill\hskip 0.3pt%
        \crcr\noalign{\kern-1pt\vskip0.07cm\nointerlineskip}%
        $\hfil\displaystyle{#1}\hfil$\crcr}}}
\def\Mc{M_{{\rm c}}}
\def\GSM{G_{{\rm SM}}}
\def\GSC{G_{{\rm SC}}}
\begin{document}
\pagestyle{empty}
\vspace{-1cm}
\noindent
\begin{flushright}
\parbox{40mm}{
KANAZAWA-01-05\\
KUNS-1720\\
NIIG-DP-01-2
}
\end{flushright}
\vspace{10mm}
\begin{center}
{\Large\bf 
Yukawa Hierarchy Transfer 
from Superconformal \\[5pt] Sector and 
Degenerate Sfermion Masses
}
\vspace*{15mm}\\
\renewcommand{\thefootnote}{\alph{footnote}}
{\large
Tatsuo {\sc Kobayashi}\rlap,\,\footnote{
E-mail: kobayash@gauge.scphys.kyoto-u.ac.jp} 
Hiroaki {\sc Nakano}\,\footnote{
E-mail: nakano@muse.hep.sc.niigata-u.ac.jp} 
and Haruhiko {\sc Terao}\,\footnote{
E-mail: terao@hep.s.kanazawa-u.ac.jp}
}
\vspace*{5mm}\\
$^a$\,Department of Physics, Kyoto University\\ 
Kyoto 606-8502, Japan 
\vspace{2mm}\\
$^b$\,Department of Physics, Niigata University\\
Niigata 950-2181, Japan
\vspace{2mm}\\
$^c$\,Institute for Theoretical Physics, Kanazawa University\\
Kanazawa 920-1192, Japan
\end{center}
\vspace*{10mm}
\begin{abstract}
We propose a new type of supersymmetric models 
coupled to superconformal field theories (SCFT's),
leading simultaneously to hierarchical Yukawa couplings
and completely degenerate sfermion masses. 
We consider models with an extra Abelian gauge symmetry
to generate hierarchical structure for couplings 
between the SM sector and the SC sector.
Interestingly, this hierarchy is inversely transferred 
to the Yukawa couplings in the SM sector.
In this type of models, 
flavor-independent structure of the superconformal fixed point
guarantees that 
the sfermion masses of the first and the second generations 
are completely degenerate at low energy.

\end{abstract}
\vspace*{20mm}
%\\
%Keywords: 

\newpage
\pagestyle{plain}
\pagenumbering{arabic}
\setcounter{footnote}{0}

The origin of the hierarchy of fermion masses and mixing angles has been
one of the most interesting problems in considering models
beyond the standard model (SM).
Especially the Yukawa texture including the lepton sector is
attracting much attention in relation with neutrino oscillation.
The most popular scenario leading to hierarchical Yukawa couplings
is the Froggatt-Nielsen (FN) mechanism \cite{FN}, in which flavor structure
is generated by an extra $U(1)_X$ gauge interaction.
The observed quark-lepton mass matrices can be explained fairly 
well by properly assigning the charges to quarks and leptons.

In general, a mechanism leading to Yukawa hierarchy
generates also flavor-dependent soft supersymmetry breaking
parameters. 
In the case of the extra $U(1)_X$ models, $D$-term contribution
to soft scalar masses becomes flavor dependent \cite{anomalous,nonanomalous}.
If $D$-term contributions are dominant, 
sfermions of the first and the second generations become 
very heavy at low energy, so that these sfermions
decouple from flavor changing processes \cite{decoupling}.
%However it has been also known that these heavy scalar masses generally
%cause color breaking problem.
With such mass spectrum of sfermions, however, one has to 
be careful to avoid color breaking problem \cite{stop}.

%Recently it has been found that degenerate sfermion masses 
%can be obtained with leading to hierarchical Yukawa couplings, 
%by considering supersymmetric standard model coupled 
%to superconformal (SC) gauge theories \cite{NS1,KT,NS2}.
Recently a new approach to the Yukawa hierarchy 
has been proposed by coupling supersymmetric standard models 
to superconformal (SC) gauge theories \cite{NS1}.
In this scenario, Yukawa couplings are exponentially suppressed 
thanks to large anomalous dimensions induced to quarks and leptons 
by an interaction with the SC sector.
The resultant mass matrices are also of Froggatt-Nielsen type.

The most intriguing property of 
models coupled with an SC sector is the suppression of 
soft supersymmetry breaking parameters \cite{NS1,KT,NS2}.
It has been found that soft scalar masses 
in a softly-broken superconformal gauge theory are subject 
to the infrared (IR) sum rules, which are given by\footnote{
Such IR behavior was first found in Ref.~\cite{KKKZ} 
for SQCD (and its dual), and subsequently 
extended to generic case in Ref.~\cite{KT}.}
\bea
\sum_i T_{\phi_i} m^2_{\phi_i} \longrightarrow 0 \ , \qquad %\nn \\
m^2_{\phi_i}+m^2_{\phi_j}+m^2_{\phi_k} \longrightarrow 0 \ .
\label{sumrule}
\eea
The first equation holds for each gauge interaction with indices $T_{\phi_i}$
for chiral superfields $\phi_i$.
The second one corresponds to each Yukawa interaction $y_{ijk}\phi_i\phi_j\phi_k$
present in the superconformal theory.
If there exists a non-renormalizable interaction 
in the superpotential at the fixed point, 
the sum of squared scalar masses for corresponding
fields is also expected to vanish \cite{KT}.
There are found  special cases that suppression of the soft masses of 
squarks and sleptons are guaranteed by these sum rules.
After the SC sector decouples, the squark and slepton masses are 
induced by the SM gaugino masses and appear nearly 
degenerate at low energy.
Thus sfermion mass degeneracy can be achieved irrespectively of the
origin of SUSY breaking.

Strictly speaking, however, the soft scalar masses are not completely 
suppressed, but converges to flavor dependent values \cite{KT,NS2}.
At the scale $\Mc$ where the SC sector decouples from the SM sector, 
the soft scalar masses of squarks and sleptons are found to converge as
\be
m^2_{q_i} 
\longrightarrow \frac{C_{ia}}{\Gamma_i}\,\alpha_a(\Mc)M^2_a(\Mc) \ ,
\label{converge}
\ee
where $\alpha_a=g_a^2/8\pi^2$ and $M_a$ denote the gauge couplings and 
the gaugino masses in the SM sector, and $C_{ia}$ denote the quadratic
Casimir.
The important point here is that
$\Gamma_i$ is a numerical factor of order $1$, 
which is obtained from the anomalous dimensions, 
and therefore is flavor dependent.
Thus the sfermion masses are fixed to flavor dependent values of
order of $\alpha M^2$. This flavor dependence is inevitable in the
Nelson-Strassler scenario, since the difference of large 
anomalous dimensions is responsible for the Yukawa hierarchy.

In Ref.~\cite{KT,NS2}, the degeneracy in low energy sfermion spectrum
is investigated by considering the SM-gaugino mass effects.
The degeneracy in the right-handed sleptons was found to be rather
weak.
As a result, the experimental bounds for lepton flavor violating
processes, {\it e.g.} $\mu \rightarrow e + \gamma$, seem difficult to be
satisfied, unless $\Gamma^{-1}_i$ is much smaller than $1$ or the 
SM gauginos are fairly heavy.

In this letter we consider a new type of models coupled to SCFTs, 
in which the hierarchical texture of fermion mass matrices and complete 
degeneracy of sfermion masses are compatible with each other.
The flavor dependence of sfermion masses given by Eq.~(\ref{converge})
arises from a difference among the 
anomalous dimensions of quarks and leptons.
Therefore we first demand that 
quarks and leptons in the first and the second generations carry 
the identical anomalous dimension at the SC fixed point.
In this case, the hierarchy of the SM-sector Yukawa couplings cannot be
generated by a difference of the large anomalous dimensions 
like in the Nelson-Strassler models, 
while the sfermion masses are degenerate completely.
In our scenario, the origin of the hierarchy resides in
the couplings between the SM sector and the SC sector,
which can be made hierarchical,
{\it e.g.}, by introducing an extra $U(1)_X$ gauge interaction.
As we will show shortly,
the SC sector plays the role not only 
of washing out flavor dependence in the soft scalar masses, 
but also of transferring the hierarchical Yukawa structure to the SM sector.

As an illustrative example, 
let us first consider a simple model based on $SU(5)$ GUT.
In the conventional FN mechanism, the flavor-dependent $U(1)_X$ charges
are sometimes assigned as $n_i=(2,1,0)$ to 
$\psi_i = (q_\L, u^c_\R, e^c_\R)_i$ in \mbox{\boldmath{$10$}},
and $-1$ to the FN field $\chi$. In that case, 
quark-lepton mass matrices are qualitatively explained
by assuming that $\chi$ develops the vacuum expectation value (VEV) 
of $\VEV{\chi} \sim M_0/20$, 
where $M_0$ is the cutoff (string) scale.

Now suppose that the chiral superfields $(\Phi_i, \widebar{\Phi}_i)$ 
in the SC sector belong to \mbox{\boldmath{$\widebar{5}$}} and 
also carry the $U(1)_X$ charges. % as well as $\psi_i$. 
Explicitly we assign $a_i$ to $\widebar{\Phi}_i\Phi_i$ 
as well as $n_i$ to $\psi_i$.
As is seen later on,
it is the $U(1)_X$ charges $a_i$ of the SC-sector fields,
not $n_i$ of SM-sector fields,
that is responsible for the desired Yukawa hierarchy.
Although $n_i$ does not play an important role in our model,
we keep it for a moment.
%So we start discussions with assuming $n_i$ to be arbitrary.
%We consider the invariant superpotential given by
The invariant superpotential is given by
\be
W = \widetilde{\lambda}_i \left(\frac{\chi}{M_0}\right)^{n_i+a_i} 
    \psi_i \widebar{\Phi}_i \Phi_i 
   +\widetilde{y}_{ij} \left(\frac{\chi}{M_0}\right)^{n_i+n_j}
    \psi_i \psi_j H \ ,
\ee
and we take the bare couplings 
$\widetilde{\lambda}_i$ and $\widetilde{y}_{ij}$
to be of order one. 
Also $H$ is the Higgs field $H$ in \mbox{\boldmath{$5$}}, 
and we assume $H$ to be $U(1)_X$ neutral for simplicity.
Since these operators are higher dimensional, the couplings 
$\widetilde{\lambda}_i$ and $\widetilde{y}_{ij}$ are 
usually thought to be suppressed towards IR.

However, we should note that the $U(1)_X$ gauge symmetry is spontaneously 
broken by non-zero VEV $\VEV{\chi}$.
After redefining the dynamical FN field at the broken vacuum, 
the superpotential also contains the $U(1)_X$ breaking terms 
\be
W = \lambda_i \psi_i \widebar{\Phi}_i \Phi_i + y_{ij} \psi_i \psi_j H \ ,
\ee
where 
$\lambda_i = \widetilde{\lambda}_i(\VEV{\chi}/M_0)^{n_i+a_i}$
and
$y_{ij} = \widetilde{y}_{ij} (\VEV{\chi}/M_0)^{n_i+n_j}$.
The operators $\psi_i \widebar{\Phi}_i \Phi_i$ are relevant thanks to
negative anomalous dimensions of the SC-sector fields.
Therefore the originally suppressed couplings $\lambda_i$ grow
and approach their IR fixed points.
Eventually the SM-sector fields $\psi_i$ acquire large anomalous dimensions,
which lead to suppression of the Yukawa couplings $y_{ij}$.
Here we restrict to the cases that the gauge group and the representations
of the SC-sector fields are identical, and therefore the anomalous dimensions
at the IR fixed points are also common for all flavors.
However, it should be noted that $\lambda_i$ reaches the fixed point at
much lower energy, if its initial value is suppressed.
As $\lambda_i$ is smaller, the growth of 
the anomalous dimension of $\psi_i$ becomes slower and therefore, 
Yukawa coupling $y_{ij}$ starts to decrease at lower energy scale.

In order to see running of the Yukawa couplings and their IR behavior
explicitly,
we examine the renormalization group equations (RGE's) for $\lambda_i$
and $y_{ij}$.
By representing the anomalous dimension of $\phi$ by $\gamma(\phi)$,
the RG equations are given by
\bea
\frac{d \ln \lambda_i}{d \ln \mu} &=&
\frac{1}{2}\leftB[ \gamma\fun{\psi_i} + 2 \gamma\fun{\Phi_i}
          \rightB] \ , \nn \\
\frac{d \ln y_{ij}}{d \ln \mu} &=&
\frac{1}{2}\leftB[ \gamma\fun{\psi_i} + \gamma\fun{\psi_j} 
                  +\gamma\fun{H}\rightB]\ . 
\eea
Here we ignored the off-diagonal elements of the anomalous dimensions,
since the diagonal ones are dominant 
due to couplings to the SC sector\rlap.\footnote{
The mixing of $\psi_i$ is induced 
by off-diagonal Yukawa couplings $y_{ij}$. 
However such effects are enough small to be neglected.
}
{}From these equations, we obtain
\be
\ln \frac{y_{ij}(M_0)}{y_{ij}(\mu)} = 
\ln \frac{\lambda_i(M_0)\lambda_j(M_0)}{\lambda_i(\mu)\lambda_j(\mu)}
+ \int_{\mu}^{M_0} \frac{d\mu'}{\mu'}
  \left[\frac{1}{2}\,\gamma(H)\fun{\mu'} - \gamma(\Phi_i)\fun{\mu'}
         -\gamma(\Phi_j)\fun{\mu'} \right] \ .
\ee
Provided that the running couplings $\lambda_i(\mu)$ approach
close to the common value of the IR fixed point,
the ratio of the Yukawa couplings are evaluated as
\bea
%\ln\frac{y_{ij}(\mu)}{y_{k\ell}(\mu)}
%&\sim & \ln\left(\frac{\VEV{\chi}}{M_0}
%           \right)^{a_k + a_\ell - a_i - a_j} \nn \\
%& & +\int_{\mu}^{M_0} \frac{d\mu'}{\mu'} 
%     \leftB[\gamma(\Phi_i)(\mu')+\gamma(\Phi_j)(\mu')
%           -\gamma(\Phi_k)(\mu')-\gamma(\Phi_\ell)(\mu')\rightB] \ ,
%
\ln\frac{y_{ij}(\mu)}{y_{k\ell}(\mu)}
\sim \ln \left(\frac{\VEV{\chi}}{M_0}\right)^{a_k+a_\ell-a_i-a_j}
    +\int_{\mu}^{M_0} \frac{d\mu'}{\mu'}
     \leftB[\gamma\fun{\Phi_i\Phi_j}\fun{\mu'}
           -\gamma\fun{\Phi_k\Phi_\ell}\fun{\mu'}\rightB] \ ,
\label{transfer} 
\eea
where we used $\lambda_i(M_0) \sim (\VEV{\chi}/M_0)^{n_i+a_i}$
as well as
$y_{ij}(M_0) \sim (\VEV{\chi}/M_0)^{n_i+n_j}$.
The difference of anomalous dimensions $\gamma(\Phi_i)$ disappears as
the SC-sector couplings approach towards the IR fixed point.
Therefore the integral in the right hand side of Eq.~(\ref{transfer})
is expected to be rather small in the situation that 
$\gamma(\Phi_i)$ are close 
to each other also at the cutoff scale $M_0$. 
Indeed the integral is found to vanish in some models 
as is explicitly shown later.

It should be noted that the ratio among SM-sector Yukawa couplings 
$y_{ij}$ are determined solely by 
the ratio among initial values $\lambda_i(M_0)$, 
{\it i.e.}, the extra $U(1)_X$ charges for the SC-sector fields $a_i$. 
In other words, the suppressed Yukawa couplings are obtained
independently of $n_i$, once all of the couplings $\lambda_i$ have
approached the IR fixed point. 
Therefore let us take $n_i=0$ hereafter.
Interestingly also the hierarchy in the SC-sector Yukawa couplings
$\lambda_i$ are transferred to the SM-sector Yukawa couplings
$y_{ij}$ in the inverse order.

The SC sector must decouple\footnote{
A naive way for decoupling is to give mass to the
SC-sector fields.
In this example, however, $\Phi_i$ and $\widebar{\Phi}_i$
cannot form their mass term. A model admitting mass terms
will be examined below.
}
from the SM sector at some scale $\Mc$,
which we assume to be common for  all SC-sector fields.
This is an important condition to
obtain the degenerate sfermion masses.
At this scale, the desired hierarchical structure 
$y_{11} \ll y_{22} \ll y_{33} \sim 1$ can be realized at
$\Mc$, if the $U(1)_X$ charges are assigned to the SC-sector fields 
so that $1 \sim \lambda_1 \gg \lambda_2 \gg \lambda_3$. Note that 
the coupling $\lambda_3$ of the third generation particles
(or top for small $\tan \beta$) to the SC sector
should be absent or kept small until the decoupling scale $\Mc$,
since their masses should not be suppressed.
The ratio $y_{11}(\Mc)/y_{22}(\Mc)$ is estimated as
$\left(\VEV{\chi}/M_0\right)^{2(a_2 - a_1)}$.

The soft scalar masses for $\psi_i$ are exponentially suppressed, 
if the IR sum rule given by Eq.~(\ref{sumrule}) can make them vanish.
This condition is equivalent to the condition that 
the anomalous dimensions of $\psi_i$ can be uniquely determined 
from the algebraic equations for anomalous dimensions 
at the fixed point \cite{KT,NS2}.
In other words, the $R$ charge of $\psi_i$ must be uniquely determined.
Therefore we seek for models such that the anomalous dimensions
for the quarks and leptons can be fixed and are identical for
different flavors.
Then the complete degeneracy of SM-sector sfermion masses as
well as Yukawa hierarchy transfer are realized.
Here we shall present several types of such models. 
\medskip \\
A. Models with $G^{N_f} \times \GSM \times U(1)_X$:

The SC-sector gauge group in the first type of models is
a product of the same groups,
$\GSC=G_1 \times G_2 \times \cdots$ with $G_i \simeq G$.
$N_f$ is the number of flavors to acquire large anomalous dimensions.
Each SC-sector field $(\Phi_i, \widebar{\Phi}_i)$ is charged 
under the $i$-th factor group $G_i$,
and is assumed to have the same representation for all $i$.
Then the anomalous dimensions $\gamma(\Phi_i)$ and 
$\gamma(\widebar{\Phi}_i)$ are uniquely determined at the fixed
point and also are flavor independent.

When we ignore the SM gauge interactions, the IR sum rules for the
soft scalar masses are given by
$m^2_{\psi_i} + m^2_{\Phi_i} + m^2_{\bar{\Phi}_i} \rightarrow 0$ and
$m^2_{\Phi_i} + m^2_{\bar{\Phi}_i} \rightarrow 0$
correspondingly to the IR attractive behavior of the SC-sector Yukawa
couplings $\lambda_i$ and SC-sector gauge couplings respectively.
Thus the soft scalar masses of $\psi_i$ are suppressed 
in this type of models.
In practice the scalar masses converge to non-zero values at IR
by the effect of the SM gauge interaction and the gaugino masses
as Eq.~(\ref{converge}).
Since these convergent values are identical, sfermions 
in the SM sector have a completely degenerate mass at low energy.
A toy model belonging to this class will be explicitly examined
later on.
\medskip \\
B. Models with non-renormalizable interactions 
$(\widebar{\Phi}_i \Phi_i)^2$:

When the anomalous dimensions of the SC fields 
$(\Phi_i, \widebar{\Phi}_i)$ are enough large in negative,
some non-renormalizable operators become relevant.
Then it is plausible that the SCFT with non-renormalizable 
interactions realizes, although the proof has not been known.
Here we assume dimension $5$ operators to appear in the superpotential
at the IR fixed point and consider the superpotential given by
\be
W = \widetilde{\lambda}_i \left(\frac{\chi}{M_0}\right)^{a_i} 
    \psi_i \widebar{\Phi}_i \Phi_i 
   +\widetilde{\zeta}_i \left(\frac{\chi}{M_0}\right)^{2a_i}
    \left(\widebar{\Phi}_i \Phi_i\right)^2+ y_{ij} \psi_i \psi_j H \ .
\ee
If $\widebar{\Phi}_i \Phi_i$ can form a singlet under
the SM gauge group, we need to forbid possible interactions
$(\chi/M_0)^{a_i}\widebar{\Phi}_i \Phi_i$ by $Z_2$ parity for example.
Otherwise SC fields decouple at the scale of the expectation value
of the FN field by obtaining masses of this order.

The breaking terms of the extra $U(1)_X$ symmetry are given by
\be
\lambda_i \psi_i \widebar{\Phi}_i \Phi_i 
+ \zeta_i\left(\widebar{\Phi}_i \Phi_i\right)^2 \ ,
\ee
with suppressed couplings $\lambda_i$ and $\zeta_i$.
If the couplings $\zeta_i$ also grow and are attracted
to a non-trivial fixed point, the anomalous dimensions of 
the SC-sector fields are fixed\footnote{
The condition for the gauge beta-function to vanish may also be satisfied
by introducing other SC-sector fields 
than $\Phi_i$ and $\widebar{\Phi}_i$.
}
as $\gamma(\Phi_i)+\gamma(\widebar{\Phi}_i) = -1$.
Therefore the anomalous dimensions $\gamma(\psi_i)$ 
of quarks and leptons are also fixed to $1$ for all flavors.

The IR sum rule for the soft scalar masses will be given by
$m^2_{\psi_i} + m^2_{\Phi_i} + m^2_{\bar{\Phi}_i} \rightarrow 0$ and
$m^2_{\Phi_i} + m^2_{\bar{\Phi}_i} \rightarrow 0$
corresponding to the IR attractive behavior of the couplings
$\lambda_i$ and $\zeta_i$ respectively.
Thus the soft scalar masses of $\psi_i$ are suppressed 
also in this type of models.
\medskip \\
C. Left-Right symmetric models with non-renormalizable interactions:

We may make models of type~B more flexible by assuming left-right
symmetry. 
For example, we introduce the gauge group
$G\times \GSM \times U(1)_X$ and chiral superfields 
$Q$ : (\mbox{\boldmath{$N$}}, \mbox{\boldmath{$\widebar{R}$}}, $0$), 
$\widebar{Q}$ : (\mbox{\boldmath{$\widebar{N}$}}, \mbox{\boldmath{$R$}}, $0$),
$P_i$ : (\mbox{\boldmath{$N$}}, {\bf 1}, $a_i$), 
$\widebar{P}_i$ : (\mbox{\boldmath{$\widebar{N}$}}, {\bf 1}, $a_i$),
$q_i$ : ({\bf 1}, \mbox{\boldmath{$R$}}, 0),  
$\widebar{q}_i$ : ({\bf 1}, \mbox{\boldmath{$\widebar{R}$}}, $0$)
and also the FN field 
$\chi$ : ({\bf 1}, {\bf 1}, $-1$).
Then the superpotential in the SC sector is written as
\be
W = \widetilde{\lambda}_i \left(\frac{\chi}{M_0}\right)^{a_i}
    q_i \widebar{P}_i Q 
   +\widebar{\widetilde{\lambda}}_i \left(\frac{\chi}{M_0}\right)^{a_i}
    \widebar{q}_i \widebar{Q} P_i
   +\widetilde{\zeta}_i \left(\frac{\chi}{M}\right)^{2a_i}
    \left(\widebar{P}_i P_i\right)^2 \ .
\ee

If these non-renormalizable interactions exist
at the fixed point, the IR sum rule tells
$m^2_{P_i} + m^2_{\bar{P}_i} \rightarrow  0$.
Since we may suppose $m^2_{P_i}= m^2_{\bar{P}_i}$ by left-right symmetry,
these masses are suppressed.
Note that
$D$-term contributions to the soft scalar masses of 
$P_i$ and $\widebar{P}_i$ are also the same 
because they carry the same $U(1)_X$ charge.
Combined with other sum rules, 
the scalar masses of $\psi_i$ are thus found to be suppressed.

Now let us examine a model of type~A explicitly by performing
numerical analysis of approximated RG equations.
Consider a model\footnote{
A prototype of this model was studied in Ref.~\cite{NS1}.
$\GSM$ is regarded as a subgroup of $E_6$,
and the quark and leptons are contained in three copies of the
$27$-dimensional representation. 
We may consider a model with $\GSM=SU(5)$ as well.
}
with $\GSC=SU(5)_1\times SU(5)_2$, 
$\GSM=SU(3)^3\times \mbox{\boldmath{$Z$}}_3$ and $U(1)_X$.
We simply assume that SM-sector fields of 
the third generation decouple from the SC sector 
and examine Yukawa couplings and sfermion masses
of the first and the second generations.
The following chiral superfields are introduced.

\begin{center}
\begin{tabular}{c|ccccc}
 & $SU(3)\times{SU(3)}\times{SU(3)}$ 
 &  $SU(5)_1$ &  $SU(5)_2$ &  $U(1)_X$ & \mbox{\boldmath{$Z$}}$_2$ 
\\[2pt] \hline
$\Phi_1$ & 
 ({\bf 3}, {\bf 1}, {\bf 1})
+({\bf 1}, {\bf 3}, {\bf 1})
+({\bf 1}, {\bf 1}, {\bf 3}) & {\bf 5} &  {\bf 1} & $0$ & $-$ 
\\[2pt]
$\widebar{\Phi}_1$ &
 ({\bf \=3}, {\bf 1}, {\bf 1})
+({\bf 1}, {\bf \=3}, {\bf 1})
+({\bf 1}, {\bf 1}, {\bf \=3}) & {\bf \=5} &  {\bf 1} & $0$ & $-$ 
\\[2pt]
$\Phi_2$ & 
 ({\bf 3}, {\bf 1}, {\bf 1})
+({\bf 1}, {\bf 3}, {\bf 1})
+({\bf 1}, {\bf 1}, {\bf 3}) & {\bf 1} &  {\bf 5} & $1/2$ & $-$ 
\\[2pt]
$\widebar{\Phi}_2$ &
 ({\bf \=3}, {\bf 1}, {\bf 1})
+({\bf 1}, {\bf \=3}, {\bf 1})
+({\bf 1}, {\bf 1}, {\bf \=3}) & {\bf 1} &  {\bf \={5}} & $1/2$ & $-$ 
\\[2pt]
$\psi_i$ &
 ({\bf 3}, {\bf \=3}, {\bf 1})
+({\bf 1}, {\bf 3}, {\bf \=3})
+({\bf \=3}, {\bf 1}, {\bf 3}) & {\bf 1} &  {\bf 1} & $0$ & $+$  
\\[2pt]
$H$ &
 ({\bf 3}, {\bf \=3}, {\bf 1})
+({\bf 1}, {\bf 3}, {\bf \=3})
+({\bf \=3}, {\bf 1}, {\bf 3}) & {\bf 1} & {\bf 1} & 0 & $+$ 
\\[2pt]
$\chi$ &  {\bf 1} &  {\bf 1} &  {\bf 1} & $-1$ & $-$ 
\end{tabular}
\end{center}

\noindent
Here \mbox{\boldmath{$Z$}}$_2$-parity is needed to forbid interactions of
$\chi \bar{\Phi}_i \Phi_i$.
We consider the superpotential %given by
\be
W = \sum_{i=1,2} \lambda_i \psi_i \widebar{\Phi}_i \Phi_i + 
\sum_{i=1,2,3} y_i \psi_i \psi_i H \ ,
\ee
where SM Yukawa matrix is restricted to diagonal one $y_i$ for
simplicity.
At the cutoff scale $M_0$, the SC-sector Yukawa couplings are
given by $\lambda_i(M_0) \sim (\VEV{\chi}/M_0)^{a_i}$ 
with $a_i =(0, 1)$.
The SM Yukawa couplings $y_i$ are also assumed to be of order 1.
In the following numerical analysis, we fix the expectation 
value to $(\VEV{\chi}/M_0)^2 = 1/300$.

The explicit form of the anomalous dimensions are known
only in perturbative expansion.
Here we use the anomalous dimensions evaluated at one-loop
order, %which are given by
\bea
\gamma(\Phi_i) = \gamma(\widebar{\Phi}_i) 
&=&\!\!\!{}-\frac{24}{5} \alpha'_i + 3 \alpha_{\lambda_i}
           -\frac{8}{3} \alpha \ , \nn \\
\gamma(\psi_i) 
&=& 5 \alpha_{\lambda_i}-\frac{16}{3} \alpha +6 \alpha_{y_i} \ .
\eea
Here $\alpha'_i = g'^2/8\pi^2$ and $\alpha = g^2/8\pi^2$
are the gauge couplings of $SU(5)_i$ and $SU(3)^3$
respectively, and 
$\alpha_{\lambda_i} = |\lambda_i|^2/8\pi^2$,
$\alpha_{y_i} = |y_i|^2/8\pi^2$.
We expect that the qualitative aspect of running couplings is 
captured by the perturbative renormalization group, 
though the beta functions are not justified near the fixed point.

First let us examine how the Yukawa hierarchy is transferred
from the SC-sector couplings $\lambda_i$ 
to the SM Yukawa couplings $y_i$ ($i=1,\,2$).
The beta functions for the couplings appearing in the superpotential
may be written down as\footnote{
In the NSVZ scheme \cite{NSVZ}, the gauge beta function becomes singular at some
strong coupling. We simply omit this pole singularity as an
approximation.
}
\bea
\mu \frac{d \alpha'_i}{d \mu} &=& \!\!\!
{}- \alpha'^2_i\leftB[6 + 9 \gamma(\Phi_i)\rightB] \ , \nn \\
\mu \frac{d \alpha_{\lambda_i}}{d \mu} &=&
\alpha_{\lambda_i}\leftB[\gamma(\psi_i)+ 2 \gamma(\Phi_i) 
                  \rightB] \ , \nn \\
\mu \frac{d \alpha}{d \mu} &=& \!\!\!
{}-\alpha^2\leftB[
-13 + 5 \gamma(\Phi_1)+ 5 \gamma(\Phi_2) 
+ 3\gamma(\psi_1)+ 3 \gamma(\psi_2)
\rightB] \ , \nn \\
\mu \frac{d \alpha_{y_i}}{d \mu} &=&
\alpha_{y_i}\leftB[2 \gamma(\psi_i) + \gamma(H)\rightB] \ ,
\label{su3^3beta}
\eea
where we neglected the anomalous dimensions 
for the third generation and Higgs fields, 
which do not couple with the SC sector.

The ratio of Yukawa couplings $y_i$ between the first and the second generations
is estimated for this model as
\be
\ln\frac{y_1(\mu)}{y_2(\mu)}
=2 \ln\frac{\VEV{\chi}}{M_0}
+2 \int_{\mu}^{M_0} \frac{d\mu'}{\mu'} 
\leftB[\gamma(\Phi_1)(\mu')-\gamma(\Phi_2)(\mu') \rightB] \ .
\label{su3^3transfer}
\ee
On the other hand the SC-sector gauge couplings are found
to be
\be
\frac{1}{\alpha'_i(\mu)} = \frac{1}{\alpha'_i(M_0)}
- 4 \ln\frac{M_0}{\mu}
- 5 \int_{\mu}^{M_0} \frac{d\mu'}{\mu'}\,\gamma(\Phi_i)(\mu') \ ,
\label{SC-gauge}
\ee
by integrating their RG equations given in 
Eqs.~(\ref{su3^3beta}).
Here note that $\alpha'_1(\mu)$ and $\alpha'_2(\mu)$ become equal
as approaching the fixed point.
If the initial gauge couplings are equal, 
$\alpha'_1(M_0)=\alpha'_2(M_0)$, 
Eq.~(\ref{SC-gauge}) tells us that
the integral in Eq.~(\ref{su3^3transfer}) vanishes
as announced before.
Therefore the Yukawa transfer is precisely realized;
$y_1(\mu)/y_2(\mu) = (\VEV{\chi}/M_0)^2$.
Even when the initial gauge couplings are different,
we find that this formula holds fairly well 
by examining numerically.

In Fig.~1, the RG running behavior of the SC-sector couplings, 
$\alpha'_i$ and  $\lambda_i$,
are shown in the case of $\alpha'_1(M_0)=\alpha'_2(M_0)=0.2$.
The running couplings of the first and the second generations
are shown by solid and dashed lines respectively.
The bold lines stands for the SC-sector gauge couplings, and
$t=\log_{10}(\mu/M)$ is the renormalization scale parameter.
Both SC-sector Yukawa couplings $\lambda_i$ approach the same value
in IR region.
The long-dashed line shows the SM-sector gauge coupling
$\alpha$, whose initial value is set to $0.2$.
An aspect of the Yukawa suppression is shown in Fig.~2 with
bare Yukawa couplings $y_i(M_0)=1.0$.
The solid and dashed lines stand for $\log_{10}y_1(\mu)$
and $\log_{10}y_2(\mu)$ respectively.

\begin{figure}[htb]
\begin{minipage}[t]{75mm}
\begin{center}
\epsfxsize=1.0\textwidth
\leavevmode
\epsffile{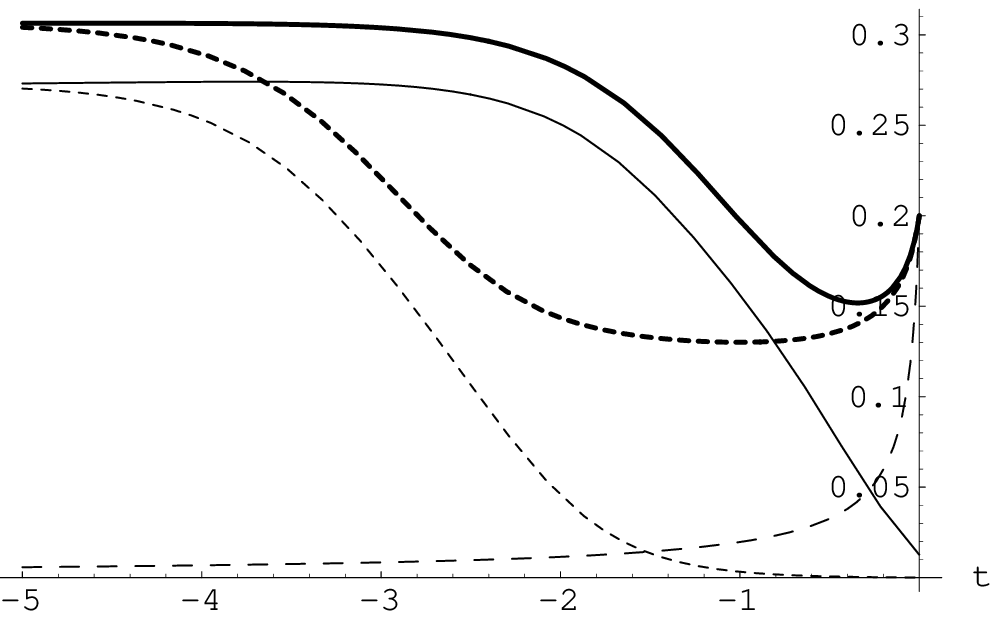}
\caption{
Running of gauge couplings, Yukawa couplings in SC-sector
and SM gauge coupling.}
\end{center}
\end{minipage}
\hspace{5mm}
\leavevmode
\begin{minipage}[t]{75mm}
\begin{center}
\epsfxsize=1.0\textwidth
\epsffile{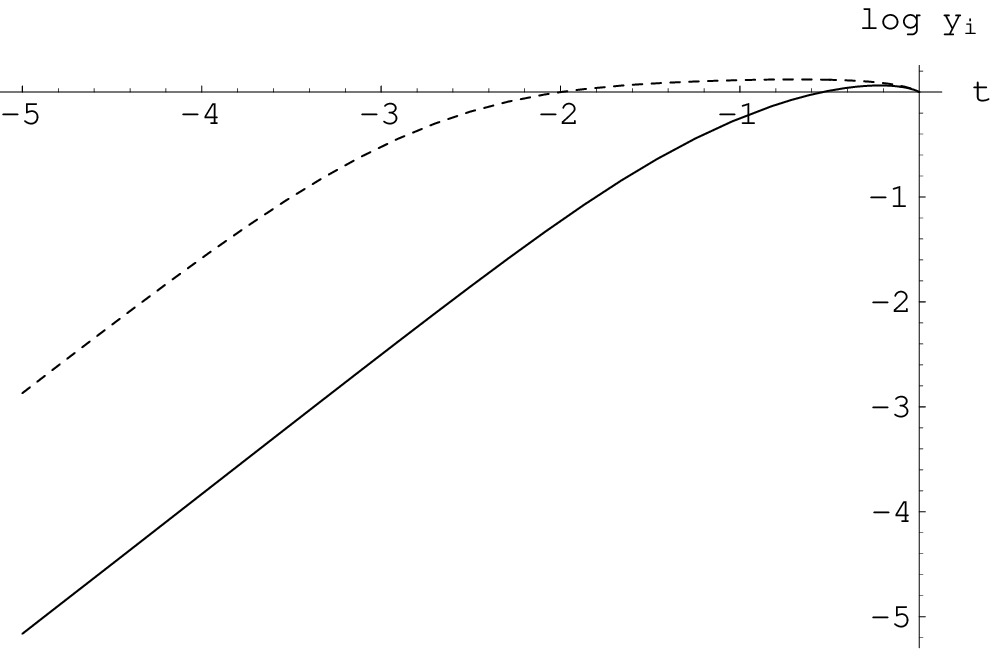}
\caption{
Power law running of Yukawa couplings and hierarchy transfer.}
\end{center}
\end{minipage}
\end{figure}

Next let us examine convergence behavior and IR degeneracy of the soft 
scalar masses.
The beta functions for the soft supersymmetry breaking parameters are
immediately derived from rigid ones by using Grasmannian expansion, 
whose explicit forms are given, for example\rlap,\footnote{
See also Refs.~\cite{softbeta}.
}
in Refs.~\cite{KT,NS2}.
In Fig.~3, the convergence behavior of soft scalar masses is
demonstrated. The bold lines and the dashed lines show 
the running behavior of soft scalar masses
$m^2_{\psi_1}$ and $m^2_{\psi_2}$ respectively 
with varying their bare parameters in $[-2.0\,m^2_0,\,2.0\,m^2_0]$.
These running couplings are obtained by analyzing 
the coupled beta functions for the soft supersymmetry breaking 
parameters deduced from Eqs.~(\ref{su3^3beta}).
The other soft parameters, gaugino masses, $A$-parameters 
and scalar masses of $(\Phi_i, \widebar{\Phi_i})$, 
are all set to $m_0$ at $M_0$.
It is seen that both soft scalar masses, 
$m^2_{\psi_1}$ and $m^2_{\psi_2}$, converge to the same value 
as the SC-sector couplings approach their fixed point.

\begin{figure}[htb]
\begin{center}
\epsfxsize=0.6\textwidth
\leavevmode
\epsffile{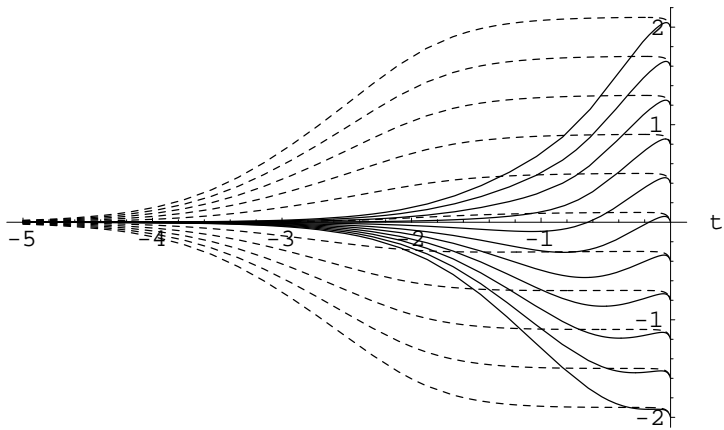}
\caption{
Convergence behavior of the soft scalar masses, 
$m^2_{\psi_1}$ and $m^2_{\psi_2}$.}
\end{center}
\end{figure}

To summarize, it is possible to construct models with 
SC sector such that the Yukawa hierarchy
and sfermion mass degeneracy are realized simultaneously.
The origin of hierarchy is attributed to the hierarchical couplings 
$\lambda_i$ of quarks and leptons to SC-sector matters, which is 
inversely transferred to the SM-sector Yukawa couplings
by superconformal dynamics.
Here we have considered models with an extra $U(1)_X$ gauge symmetry
to make these couplings $\lambda_i$ hierarchical, although the origin 
of the hierarchy may be introduced in some other ways.
The flavor independent structure of the superconformal fixed point 
ensures the degenerate sfermion masses at low energy.

We may consider other scenarios using superconformal field theories
to wash out the flavor dependence in the sfermion masses.
For example, it is possible to eliminate the $D$-term contributions 
in the sfermion masses.
It has been known that the $D$-term contributions are proportional 
to the scalar mass of the FN field~\cite{anomalous}.
Therefore, if this scalar mass of the FN field is suppressed 
by proper coupling with an SC sector, 
then the $D$-term contributions are washed out.
Such scenarios will be reported elsewhere \cite{KNNT}.

\end{document}